\documentclass[twocolumn]{article}

\usepackage[margin=0.75in]{geometry}   

\usepackage{longtable}
\usepackage{subcaption}  
\usepackage{amsmath,amssymb}
\usepackage{graphicx}
\usepackage{booktabs}
\usepackage{siunitx}
\usepackage{threeparttable}
\usepackage{tabularx}                  
\usepackage{caption}
\usepackage{natbib}
\usepackage{url}
\usepackage{hyperref}
\usepackage{enumitem}
\usepackage{placeins}
\usepackage{float}

\newcolumntype{L}{>{\raggedright\arraybackslash}X} 

\title{\bfseries Borrowing on Belief? Consumer Confidence and U.S.\ Credit—A VECM Study}

\author{
  \ Samiha Tariq \ Weikang Zhang\\[0.5ex]
  \  Southern Illinois University Carbondale
}
\date{May 2025}

\begin{document}
\maketitle

\begin{abstract}
\itshape
This study explores the interdependent relationship between consumer credit and consumer confidence in the United States using monthly data from January~1978 to August~2024. Utilizing a Vector Error Correction Model (VECM), the analysis focuses on the interplay between household borrowing behaviour and consumer sentiment while \emph{controlling} for macroeconomic factors such as interest rates, inflation, unemployment, and money supply. The results reveal a stable long‑run equilibrium: heightened consumer confidence is associated with increased credit utilisation, reflecting greater financial optimism among households. In the short run, shifts in consumer confidence exert relatively modest immediate influence on credit usage, whereas consumer credit adjusts slowly, displaying significant inertia. Impulse‑response analysis confirms that shocks to consumer confidence generate sustained positive effects on borrowing, while unexpected increases in credit initially depress sentiment but only fleetingly. These findings underscore the critical role of the relationship between consumer confidence and credit‑market dynamics and highlight its policy relevance for fostering balanced and stable household finances.
\end{abstract}

\section{Introduction}

Consumer credit remains an essential indicator of household financial behaviour in the United States, and the past five years have seen that barometer swing sharply in step with the pandemic, inflation, and monetary-policy reversals. Seasonally adjusted Federal Reserve data show that total outstanding consumer credit reached \$5.01 trillion in March 2025, up roughly \$570 billion from its pandemic trough and continuing the gentle up-trend evident since mid-2021 (Federal Reserve Board, 2025; YCharts, 2025). Revolving balances have accounted for a growing share of that expansion. Credit-card debt fell from its pre-COVID peak of \$927 billion in 2019 Q4 to \$770 billion in 2021 Q1 as households deleveraged in lockdown, then surged to a record \$1.21 trillion by 2024 Q4—about 7 percent higher than a year earlier—amid rising prices and tighter budgets (Federal Reserve Bank of New York, 2020, 2024).

Consumer sentiment has moved in the opposite direction. The University of Michigan index slid from 101.0 in February 2020 to 71.8 just two months later and, despite intermittent rebounds, slipped to 52.2 in April 2025 and an even weaker 50.8 in the preliminary May reading, underscoring renewed worries about inflation and growth (University of Michigan, 2020; Reuters, 2025).

Those shifts in borrowing and confidence unfolded against a dramatic policy backdrop. The Federal Open Market Committee cut the target federal-funds range to 0–0.25 percent on 15 March 2020, then embarked on its fastest hiking cycle in four decades, taking the range to 5.25–5.50 percent by July 2023 before easing to 4.25–4.50 percent in March 2025 as inflation cooled (Federal Reserve, 2020, 2025). Labour-market damage from COVID-19 was equally stark: unemployment spiked to 14.7 percent in April 2020—the highest recorded since the 1930s—yet fell back to 4.2 percent by April 2025, helping to stabilise household incomes even as borrowing costs rose (Bureau of Labor Statistics, 2020, 2025).

This paper examines the long-run and short-run dynamics between consumer confidence and consumer credit. Understanding these linkages is critical for assessing household resilience and channels for policy-making.

\section{Literature Review}

The existing literature has examined the relationship between consumer credit and consumer confidence across various contexts, highlighting diverse perspectives and methodologies. Su, Peculea, and Wang (2023) examine the dynamic interplay between consumer credit (CC) and consumer confidence (CF) in China through a bootstrap rolling-window causality test. Their findings suggest that consumer credit can positively influence consumer confidence during certain periods by alleviating liquidity constraints and temporarily enhancing purchasing power, thereby improving financial flexibility. However, the study also reveals negative relationships during other periods, primarily due to policy adjustments and financial instabilities that can erode consumer confidence, particularly when individuals become cautious about future economic conditions. Additionally, consumer confidence is shown to negatively impact consumer credit under specific circumstances, as shifts in consumer expectations about future economic situations alter borrowing behaviours. These interactions are further complicated by external factors, including governmental interventions and unexpected events such as the COVID-19 pandemic, disrupting typical credit-confidence dynamics.

Douglas~J.~Lamdin (2008) investigates the relationship between consumer sentiment and revolving credit use, with a particular emphasis on U.S.\ credit-card debt. Over three decades, revolving credit usage has surged significantly, increasing by over 600\% in inflation-adjusted terms. Lamdin’s analysis explores whether shifts in the University of Michigan’s Index of Consumer Sentiment can predict future variations in revolving credit use. The study identifies a positive correlation between consumer sentiment changes and subsequent revolving credit usage, indicating that consumer confidence indeed serves as a reliable predictor of future credit behaviours.

An, Cordell, Roman, and Zhang (2023) explore the impact of Federal Reserve monetary policies—such as interest-rate adjustments and quantitative easing—on household financial activities, including borrowing, refinancing, and home purchasing. Their findings suggest a diminished transmission of lower mortgage rates during the COVID-19 pandemic, primarily due to the emergence of shadow-bank mortgage lenders who, despite charging higher fees and rates, offered faster and more convenient loan processing. This phenomenon facilitated timely monetary-policy implementation during crisis periods. These monetary-policy shifts also significantly influence consumer confidence, subsequently affecting broader economic activities through spending and investment patterns.

Başarır (2022) explores the interactions among credit-card expenditures, consumer confidence, and saving behaviours between March~2014 and December~2021 using the ARDL model. The study concludes that long-term credit-card spending negatively impacts consumer confidence, whereas saving tendencies do not significantly affect it. In the short run, however, saving behaviours notably influence consumer confidence, and lagged credit-card expenditures have a slight positive impact. This research underscores the complexity of consumer spending and saving behaviours, especially amid uncertainties like the COVID-19 pandemic, in shaping economic sentiment and decision-making.

Gündüz (2017) investigates the relationship between consumer confidence and household credit-card expenditures in Turkey, using monthly data spanning January~2004 to January~2006. The study assesses how consumer confidence, indicative of future financial and economic expectations, affects spending behaviours. The findings from regression and Granger-causality analyses confirm a unidirectional causal relationship from consumer confidence to household credit-card expenditures. Moreover, increases in consumer confidence from previous periods consistently lead to higher subsequent credit-card spending.

Han et~al.\ (2022) analyse both short- and long-term impacts of credit accessibility on consumption through a quasi-experimental design and detailed transaction data. They find that immediate consumption significantly rises by 51.74\% following credit access but declines by 4.02\% over the longer term as consumers adjust to financial constraints. Additionally, the study identifies spill-over effects, illustrating how credit usage influences consumption indirectly through savings. Regulatory-focus theory is employed to interpret these behavioural shifts, providing crucial insights for financial-marketing strategies and policy design.

James~P.~Dow (2018) employs data from the U.S.\ Survey of Consumer Finances to examine changing attitudes toward credit post-Great Recession. The research identifies considerable heterogeneity in credit attitudes among households, although overall sentiment changes post-crisis are moderate, with most households becoming marginally more conservative in credit usage. Demographic factors—including age, race, and gender—significantly shape specific credit attitudes, highlighting the nuanced ways in which demographic characteristics influence credit perceptions.

The present study extends the existing literature by broadening the analytical scope and depth through the use of extensive monthly data from FRED, covering January~1978 to August~2024. This comprehensive dataset captures multiple economic cycles and significant events, including the 2008 financial crisis and the COVID-19 pandemic, enabling a robust analysis of consumer-credit behaviour trends. Unlike previous studies constrained by fewer variables or shorter timeframes, this research incorporates consumer confidence, interest rates, inflation, money supply (M2), and unemployment rates. Such integration provides a nuanced understanding of consumer-credit interactions with broader macroeconomic conditions, distinguishing between psychological factors such as consumer confidence and economic fundamentals like interest rates and M2. By addressing gaps in earlier research that typically emphasised short-term impacts or limited geographic and temporal scopes, this study significantly contributes to the literature by elucidating the dynamic relationships among consumer credit, consumer confidence, and macroeconomic factors over an extended period.

This investigation draws upon Behavioural Economics—specifically Prospect Theory—to explore the relationship between consumer credit and consumer confidence within monetary economics. Traditional economic models assume rational decision-making based on objective criteria such as interest rates, income, and liquidity constraints. In contrast, behavioural economics highlights psychological and emotional factors that often lead to departures from rationality. Prospect Theory, developed by Kahneman and Tversky (2013), describes how individuals assess potential outcomes relative to a reference point rather than in absolute terms, emphasising behavioural patterns such as loss aversion, framing effects, and varying risk preferences. These patterns significantly influence borrowing behaviours, particularly through the mediating role of consumer confidence, thereby enriching the understanding of consumer-credit dynamics and offering valuable implications for monetary-policy design and implementation.


\section{Data \& Research Methodology}

This study employs monthly secondary data sourced from the Federal Reserve Economic Data (FRED) database, spanning January 1978 through August 2024. This extended period offers sufficient observations to comprehensively analyze both short-term dynamics and long-term equilibrium relationships between consumer credit and key economic indicators, with particular emphasis on the interplay between consumer credit and consumer confidence.

The dependent variable of interest is \emph{total consumer credit}, measured in trillions of U.S.~dollars and encompassing all consumer credit owned and securitized. This aggregate captures both revolving credit—such as credit-card debt—and non-revolving credit, including auto and personal loans, thereby providing a holistic indicator of household borrowing behaviour.

\emph{Consumer confidence}, proxied by the University of Michigan Consumer Sentiment Index (UMCSENT), captures the psychological dimension of household behaviour, reflecting optimism or pessimism regarding both the broader economic outlook and personal financial conditions. This variable is fundamental because perceptions of economic security directly influence borrowing decisions.

To account for monetary-policy effects, the Federal Funds Rate (FEDFUNDS) is incorporated. As the primary instrument of U.S.~monetary policy, variations in this rate materially affect borrowing costs and, by extension, consumer credit demand. \emph{Inflation}, measured by monthly percentage changes in the Consumer Price Index (CPIAUCSL), captures fluctuations in purchasing power that may compel consumers to adjust borrowing to maintain consumption levels during periods of rising prices.

The \emph{unemployment rate} (UNRATE) is included because employment stability conditions consumers’ capacity to service debt. Additionally, the broad measure of money supply (M2SL) gauges liquidity conditions in the economy, reflecting overall credit availability.

To model potential long-run equilibrium relationships among these variables, the analysis adopts a Vector Error Correction Model (VECM), which is well suited for cointegrated time-series data. The VECM captures both short-term adjustments and long-run equilibrium dynamics. Its general specification is

\begin{equation}
\Delta \mathbf{Y}_{t}
  = \Pi \mathbf{Y}_{t-1}
    + \sum_{i=1}^{k-1} \Gamma_{i}\,\Delta \mathbf{Y}_{t-i}
    + \boldsymbol{\Phi}\mathbf{X}_{t}
    + \boldsymbol{\epsilon}_{t},
\end{equation}

where $\Delta\mathbf{Y}_{t}$ is the vector of first-differenced endogenous variables (consumer credit, consumer confidence, interest rates, inflation, unemployment, and money supply); $\Pi$ captures long-run relationships and adjustment speeds; the $\Gamma_{i}$ matrices model short-term dynamics; $\boldsymbol{\Phi}\mathbf{X}_{t}$ collects deterministic terms such as intercepts or trends; and $\boldsymbol{\epsilon}_{t}$ denotes independently and identically distributed error terms. This framework permits a nuanced exploration of consumer-credit dynamics, with a focus on how consumer confidence shapes household borrowing over time.

Summary statistics (Appendix: Table 1) reveal pronounced upward trends in credit and money supply and volatility in sentiment.  Augmented Dickey--Fuller (Appendix: Table 2) tests confirm that credit, consumer sentiment, CPI, and M2 are non‑stationary, while the federal‑funds and unemployment rates are stationary after first differencing.

\begin{figure}[htbp]        
  \centering
  \includegraphics[width=0.8\columnwidth]{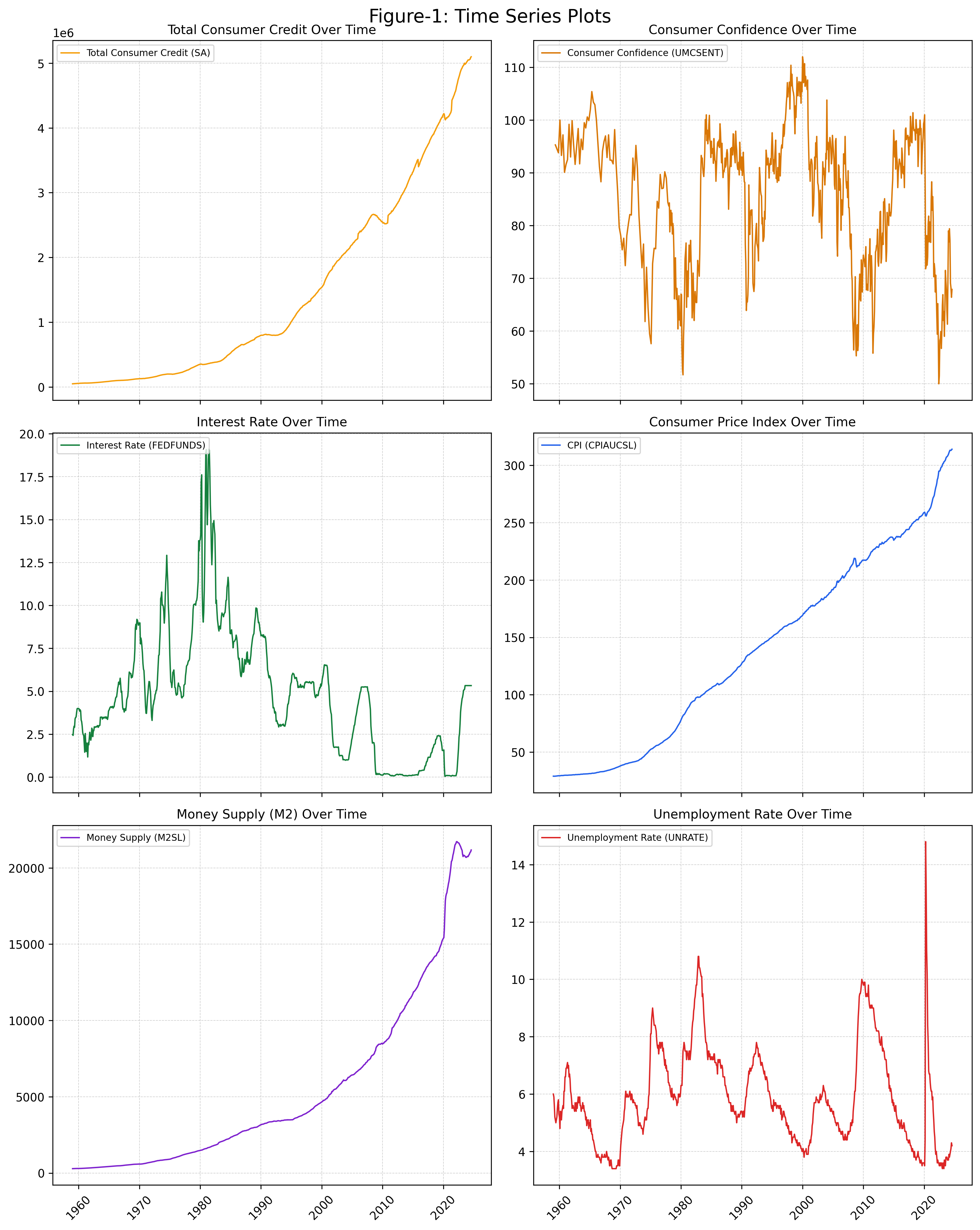}

  \caption{Time–series evolution of key U.S.\ macro-financial indicators,
           1960–2024: total consumer credit, consumer confidence, federal
           funds rate, CPI, money supply (M2), and unemployment rate.}
  \label{fig:timeseries}
\end{figure}

Figure 1 shows two key patterns. First, total consumer credit follows a near-exponential upward path. Second, consumer confidence oscillates sharply across business cycles. Interest rates, inflation, money supply, and unemployment are plotted for contextual control—they affect borrowing costs, real purchasing power, and labour-market security—but the empirical focus remains on how the sustained expansion of credit aligns with the cyclical swings in sentiment.

\section{Empirical Analysis}

\subsection{Lag-Order Selection and Johansen Cointegration Test}
The optimal lag length was chosen with the Schwarz Information Criterion
(SIC) and the Hannan–Quinn Information Criterion (HQIC); both point to two
lags as the best trade-off between parsimony and fit.  
Accordingly, the Johansen
trace test is estimated with \(p = 2\) lags.

The trace statistics (Appendix: Table 3) decisively reject the null
hypotheses of no cointegration (\(r \le 0\)), at most one cointegrating
relationship (\(r \le 1\)), and at most two cointegrating relationships
(\(r \le 2\)) at the \(5\%\) level.  The corresponding trace values
\(143.462\), \(72.958\), and \(30.582\) all exceed their critical values
\(69.819\), \(47.855\), and \(29.796\), respectively.
Although the test even rejects \(r \le 4\), implying full rank,
this contradicts the unit-root nature of the individual series
(\(I(1)\)).  Balancing statistical evidence with economic
interpretability, a cointegration rank of two is imposed in the subsequent
analysis.

\subsection{Vector Error-Correction Model (VECM)}

\subsubsection{Long-Run Equilibrium Relationships}
Our analysis (Appendix: Table 10) reports the two normalised cointegrating vectors.
The first vector shows that higher consumer confidence raises total consumer
credit, whereas tighter monetary conditions (federal funds rate) and broad
monetary aggregates exert downward pressure.  A strong positive CPI
coefficient confirms the long-run inflation effect on credit.

The second vector indicates that consumer confidence itself is positively
driven by total credit, negatively affected by the federal funds rate, and
positively related to inflation (CPI) and the money supply (M2).

\subsubsection{Short-Run Dynamics}
Estimated short-run coefficients for the total-credit equation
(Appendix: Table 4) reveal pronounced inertia:
the first and second lags of credit enter with positive and highly
significant coefficients (\(0.127\) and \(0.107\), both \(p<0.01\)).
Lagged consumer-confidence terms are insignificant, while money supply (M2)
has a mixed short-run impact—negative in the first lag
(\(-32.225\), \(p<0.01\)) and positive in the second
(\(41.256\), \(p<0.01\))—reflecting complex liquidity effects.

Consumer confidence (Appendix: Table 5) is more sensitive to
contemporaneous macro factors.  Significant drivers include its own first
lag (\(-0.1034\), \(p<0.05\)), the second lag of the federal funds rate
(\(-0.8939\), \(p<0.01\)), the first lag of CPI
(\(-1.0467\), \(p<0.01\)), and the first (\(-0.0095\), \(p<0.05\)) and
second (\(0.0063\), \(p<0.10\)) lags of M2.

\subsubsection{Error-Correction Mechanism}
The adjustment coefficients confirm gradual convergence to the long-run
equilibrium.  In the credit equation, the error-correction term associated
with the first cointegrating vector is \(-0.001\) (\(p<0.01\)),
implying a slow adjustment speed of roughly \(0.1\%\) per month.
Consumer confidence corrects disequilibrium even more slowly
(\(\mathrm{EC}_{1,t-1} = -3.3\times10^{-7}\), \(p<0.05\)).

\subsection{Summary of Empirical Findings}

\noindent\textbf{Long‐Run Complementarity:} A strong positive equilibrium link exists between consumer confidence and consumer credit, in line with theories that optimism fuels borrowing. \textbf{Monetary Forces Dominate Short‐Run Dynamics:} Short-term credit movements respond mainly to liquidity (M2) rather than transient sentiment shifts. \textbf{Gradual Equilibrium Restoration:} Adjustments toward long-run balance are slow for both credit and confidence, underscoring the influence of persistent macro forces over abrupt corrections. 

\medskip

\noindent\textit{Policy implication:} Monetary stimulus that bolsters consumer confidence can amplify its effectiveness, given their tight long-run relationship.

\subsection{Impulse--response functions}\label{subsec:irf}

The impulse–response functions (IRFs) displayed in Figure 2 reveal the dynamic feedback between consumer confidence and consumer credit implied by the VECM.
\begin{figure}[htbp]  
  \centering

  \begin{subfigure}{0.95\columnwidth}  
    \centering
    \includegraphics[width=\linewidth]{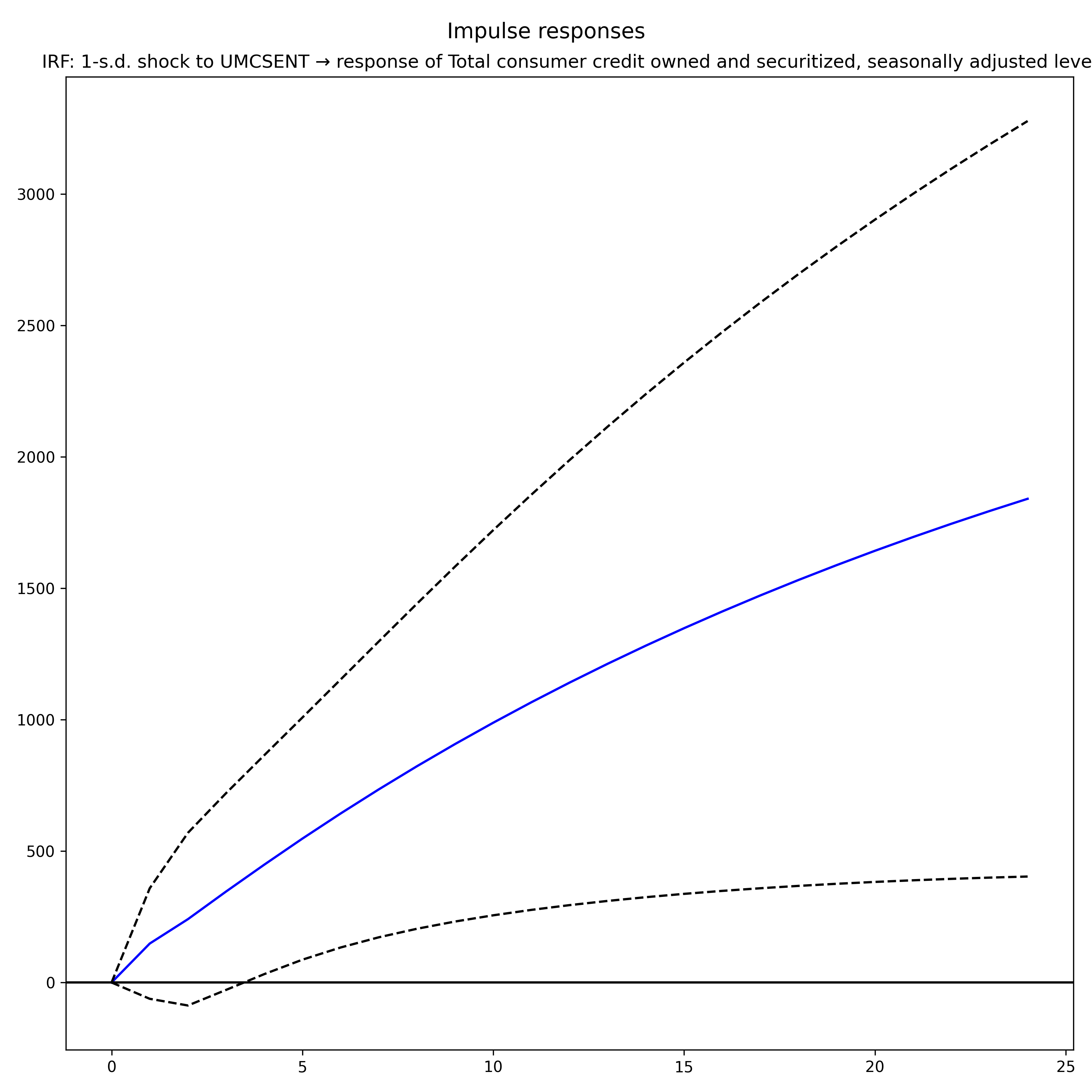}
    \caption{Response of total consumer credit to a one–s.d.\ consumer-confidence shock.}
    \label{fig:irf_credit_from_conf}
  \end{subfigure}

  \vspace{0.8em} 

  \begin{subfigure}{0.95\columnwidth}
    \centering
    \includegraphics[width=\linewidth]{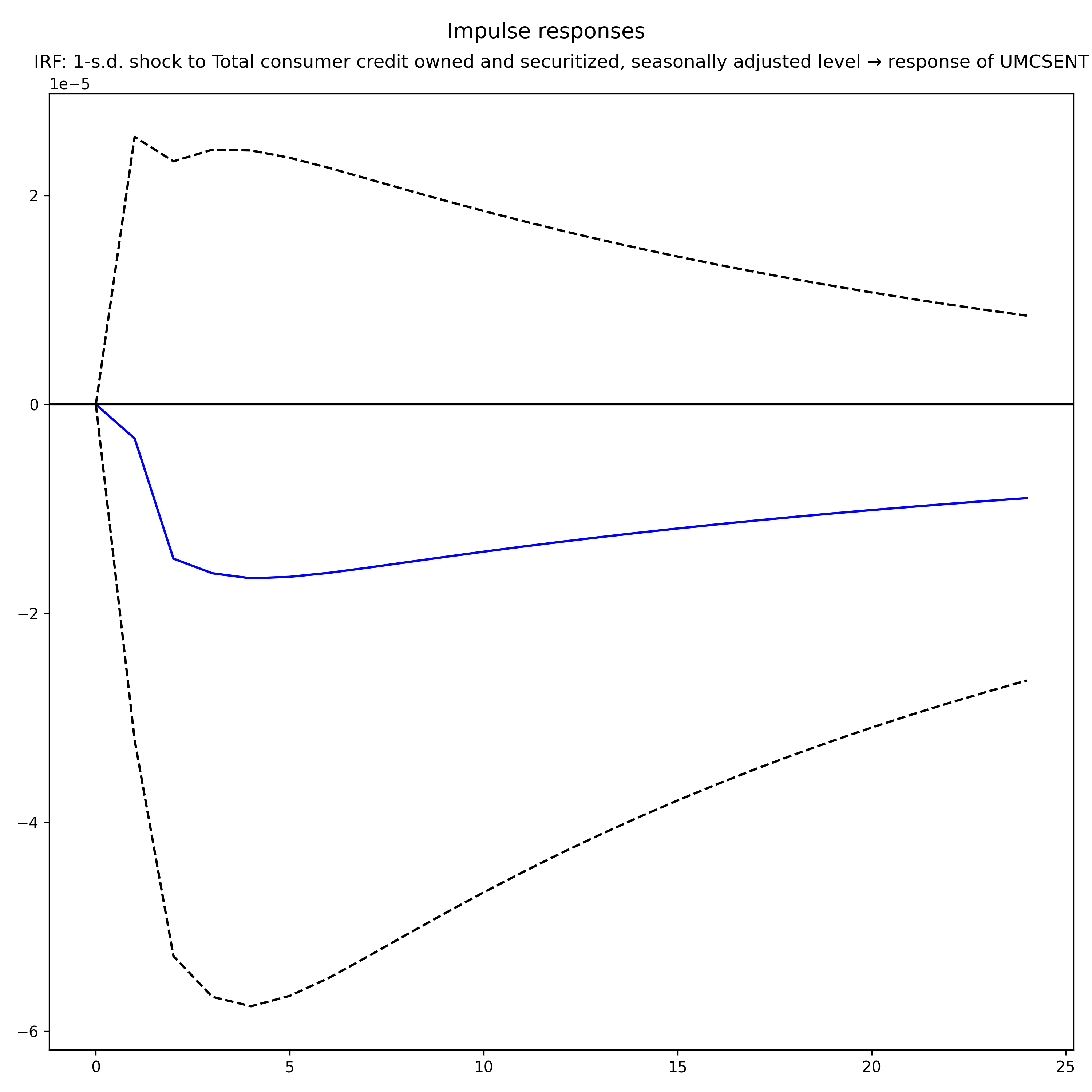}
    \caption{Response of consumer confidence to a one–s.d.\ credit shock.}
    \label{fig:irf_conf_from_credit}
  \end{subfigure}

  \caption{Impulse–response functions from the VECM, with 95\% confidence bands (dashed).}
  \label{fig:irf_pair}
\end{figure}

\paragraph{Credit response to a confidence shock.}
Panel (a) of Figure 2 traces the effect of a one–standard–deviation (\(1\sigma\)) innovation in \emph{consumer confidence} on \emph{total consumer credit}.  The response is positive from the outset, reflecting the lift that heightened optimism gives to borrowing.  Over subsequent periods this effect compounds, generating a steadily rising credit path.  Although the confidence band widens with the horizon—signalling greater uncertainty about the exact magnitude—the central trajectory remains firmly upward, underscoring a strong and persistent confidence–credit transmission mechanism.

\paragraph{Confidence response to a credit shock.}
Panel (b) of Figure 2 shows the reverse experiment: the reaction of \emph{consumer confidence} to a \(1\sigma\) disturbance in \emph{total consumer credit}.  The immediate impact is a modest dip, suggesting that a sudden build-up in household debt initially raises concerns over financial sustainability.  This adverse effect, however, fades gradually.  The impulse path edges back toward the baseline while the confidence band narrows, indicating that sentiment normalises in the medium to long run.

Taken together, these IRFs confirm a stable long-run equilibrium between the two variables.  Consumer confidence persistently drives credit expansion, whereas surges in credit impose only temporary headwinds on confidence.  This nuanced two-way interaction has clear implications for the calibration of monetary and fiscal policies aimed at sustaining household balance-sheet resilience while fostering economic optimism.

\section{Conclusion}
This study examined the intricate relationship between consumer credit and consumer confidence in the United States using a Vector Error Correction Model (VECM), analyzing monthly data from 1978 to 2024. The empirical results indicate the presence of two stable long-run relationships, underscoring the deep structural linkage between household borrowing behavior and consumer sentiment. Specifically, higher levels of consumer confidence significantly encourage greater credit utilization in the long run, while tighter monetary conditions and rising inflationary pressures tend to restrain borrowing.
The short-run dynamics further reveal nuanced interactions: changes in money supply predominantly drive short-term fluctuations in consumer credit, whereas adjustments in consumer confidence occur more gradually in response to macroeconomic shocks. Impulse response analyses further illustrate these dynamics clearly. Positive shocks to consumer confidence generate persistent increases in credit, reinforcing the vital role of consumer optimism in stimulating sustained credit expansion. Conversely, unexpected expansions in consumer credit initially moderate consumer sentiment, reflecting cautious household responses to increased debt burdens, although this impact tends to diminish relatively quickly.
These findings provide important insights into the psychological underpinnings of consumer financial behavior, highlighting how consumer confidence not only mirrors economic conditions but actively shapes borrowing decisions. For policymakers, this emphasizes the necessity of policies that jointly address economic fundamentals and consumer perceptions. Initiatives fostering consumer optimism, such as credible monetary frameworks and clear communication strategies, can meaningfully enhance the effectiveness and durability of broader economic stabilization efforts.
While certain methodological and data-specific factors may introduce complexity, the robust empirical evidence presented in this study strongly affirms the central role of consumer confidence in shaping the landscape of consumer credit. Ultimately, this research underscores that credit dynamics in the U.S. are fundamentally linked to consumer sentiment—highlighting how optimism and confidence serve as critical drivers of household financial behavior and broader macroeconomic stability.

\section{References}
An,~X., Cordell,~L., Roman,~R.~A., \& Zhang,~C.\ (2023). Central bank monetary policy and consumer credit markets. \textit{Oxford Research Encyclopedia of Economics and Finance.} https://doi.org/10.1093/acrefore/9780190625979.013.XXX

Başarır,~Y.\ (2022). The relationship between credit card expenditures, consumer confidence and consumers’ saving tendencies. \textit{Journal of Empirical Economics and Social Sciences}, 4(1), 65--77. https://doi.org/10.46959/jeess.1080739

Bureau of Labor Statistics.\ (2020, May 13). Unemployment rate rises to record high 14.7 percent in April~2020. https://www.bls.gov/opub/ted/2020/unemployment-rate-rises-to-record-high-14-point-7-percent-in-april-2020.htm

Bureau of Labor Statistics.\ (2025, May 3). The employment situation---April 2025. https://www.bls.gov/news.release/pdf/empsit.pdf

Dow,~J.~P.\ (2018). Attitudes towards credit after the Great Recession. \textit{Applied Economics Letters}, 25(4), 254--257. https://doi.org/10.1080/13504851.2017.1321836

Federal Reserve Bank of New York.\ (2020). textit{Quarterly report on household debt and credit: 2019 Q4}. 

Federal Reserve Bank of New York.\ (2024). \textit{Household debt and credit report: 2024 Q4}.

Federal Reserve.\ (2020, March 15).FOMC statement. https://www.federalreserve.gov/monetarypolicy/fomcminutes20200315.htm

Federal Reserve.\ (2025, March 19). FOMC statement. https://www.federalreserve.gov/newsevents/pressreleases/monetary20250319a.htm

Federal Reserve Board.\ (2025, May 7). Consumer credit---G.19 statistical release. https://www.federalreserve.gov/releases/g19/current/g19.pdf

Gündüz,~İ.~O., Sönmezler,~G., \& Akduğan,~U.\ (2017). An analysis of the relationship between consumer confidence index and credit card expenditures in Turkey. textit{IIB International Refereed Academic Social Sciences Journal, 8}(27), 1--16.

Han,~J., Zhang,~Y., Lu,~T., Sun,~Y., \& Huang,~W.\ (2022). Brake or step on the gas? Empirical analyses of credit effects on individual consumption. \textit{Economic Modelling, 118}, 105985. https://doi.org/10.1016/j.econmod.2022.105985

Lamdin,~D.~J.\ (2008). Does consumer sentiment foretell revolving credit use? \textit{Journal of Family and Economic Issues, 29}(2), 279--288. https://doi.org/10.1007/s10834-008-9108-9

Reuters.\ (2025, May 16). US consumer sentiment slumps, households brace for inflation surge.

Su,~C.-W., Peculea,~A.~D., \& Wang,~K.-H.\ (2023). Can consumer credit stimulate consumer confidence? Evidence from the time‑varying aspect. \textit{Annals of Financial Economics, 18}(2), 2350017. https://doi.org/10.1142/S0217590823500170

University of Michigan.\ (2020). \textit{Monthly data tables---May 2020}.

YCharts.\ (2025). \textit{US total consumer credit outstanding}.\\
\url{https://ycharts.com/indicators/total_consumer_credit_outstanding}

\FloatBarrier   
\clearpage
\onecolumn
\appendix
\section*{Appendix}
\FloatBarrier   

\begin{table}[H]   
  \centering\small  
  \begin{threeparttable}
    \caption{Descriptive Statistics}
    \label{tab:descr_stats}
    \setlength{\tabcolsep}{5pt} 
    \begin{tabular}{
      l
      S[table-format=3.0]      
      S[table-format=8.3]      
      S[table-format=8.3]      
      S[table-format=7.3]      
      S[table-format=9.3]      
    }
      \toprule
      {Variable} & {Observ.} & {Mean} & {Std.\ Dev.} & {Min} & {Max} \\
      \midrule
      Total consumer credit & 788 & 1454936.979 & 1452142.626 & 48961.160 & 5097589.770 \\
      Consumer Confidence   & 634 &      85.217 &      13.037 &     50.000 &      112.000 \\
      Interest Rate         & 788 &       4.777 &       3.630 &      0.050 &       19.100 \\
      Inflation             & 788 &     137.997 &      82.921 &     28.970 &      314.121 \\
      M2 Money Supply       & 788 &    5346.139 &    5667.009 &    286.600 &    21722.000 \\
      Unemployment Rate     & 788 &       5.890 &       1.685 &      3.400 &       14.800 \\
      \bottomrule
    \end{tabular}
    \begin{tablenotes}[flushleft]
      \footnotesize
      \item \textit{Notes}: Monthly data, January 1978–August 2024. Monetary figures in millions of dollars.
    \end{tablenotes}
  \end{threeparttable}
\end{table}

\begin{table}[htbp]        
  \small
  \centering
  \caption{Augmented Dickey--Fuller (ADF) Unit-Root Test Results}
  \label{tab:ADF}

  \begin{threeparttable}
  \begin{tabularx}{\textwidth}{
    L
    S[table-format = 2.3]      
    S[table-format = 1.4]      
    S[table-format = -1.3]     
    S[table-format = -1.3]     
    l                          
  }
    \toprule
    \textbf{Variable} &
    {\textbf{ADF}} &
    {\textbf{$p$-value}} &
    {\textbf{CV 1\%}} &
    {\textbf{CV 5\%}} &
    \textbf{Stationarity} \\
    \midrule

    Total consumer credit owned and securitized, \textit{not} seasonally adjusted level &
      5.022 & 1.0000 & -3.439 & -2.865 & Not stationary \\

    UMCSENT (Consumer sentiment index) &
      -2.837 & 0.0532 & -3.441 & -2.866 & Not stationary \\

    FEDFUNDS &
      -2.907 & 0.0445 & -3.439 & -2.865 & Stationary \\

    CPIAUCSL (CPI) &
      1.900 & 0.9985 & -3.439 & -2.865 & Not stationary \\

    M2SL (Money supply) &
      3.725 & 1.0000 & -3.439 & -2.865 & Not stationary \\

    UNRATE (Unemployment rate) &
      -3.422 & 0.0102 & -3.439 & -2.865 & Stationary \\
    \bottomrule
  \end{tabularx}

  \begin{tablenotes}[flushleft]
    \footnotesize
    \item \textit{Note.} $H_{0}$: the series has a unit root (non-stationary).
    “CV 1\%” and “CV 5\%” are the critical values at the 1 \% and 5 \% significance levels.
    We reject $H_{0}$ when the ADF statistic is smaller than the critical value and/or the $p$-value $<0.05$.
  \end{tablenotes}
  \end{threeparttable}
\end{table}                 

\begin{table}[H]
  \centering
  \caption{Johansen Trace Test (lag difference = 2)}
  \label{tab:trace}
  \begin{threeparttable}
    \begin{tabular}{
      l
      S[table-format=6.3]
      S[table-format=6.3]
    }
    \toprule
    {Null Hypothesis} & {Trace statistic} & {95\% critical value} \\
    \midrule
    $r \le 0$ & 143.462 & 69.819 \\
    $r \le 1$ &  72.958 & 47.855 \\
    $r \le 2$ &  30.582 & 29.796 \\
    $r \le 3$ &  17.220 & 15.494 \\
    $r \le 4$ &   5.777 &  3.841 \\
    \bottomrule
    \end{tabular}
  \end{threeparttable}
\end{table}

\newcommand{\LagTableHeader}{
  \toprule
  {Regressor} &
  {Coef.} & {Std.\ Err.} & {$z$} & {$P>|z|$} &
  {95\% CI\,Lower} & {95\% CI\,Upper}\\
  \midrule
}

\begin{table}[H]
  \centering\scriptsize
  \caption{Lagged Parameters for Equation: Total Consumer Credit (Level)}
  \label{tab:tcc}
  \begin{tabularx}{\textwidth}{l*6{S[table-format=8.4]}}
  \LagTableHeader
  L1.Total consumer credit & 0.1271 & 0.0420 & 3.054 & 0.002 & 0.046 & 0.209 \\
  L1.UMCSENT              & -3.5859 & 109.0100 & -0.033 & 0.974 & -217.242 & 210.070 \\
  L1.FEDFUNDS             & 458.3989 & 788.8450 & 0.581 & 0.561 & -1087.709 & 2004.506 \\
  L1.CPIAUCSL             & 1635.2709 & 1021.6020 & 1.601 & 0.109 & -367.033 & 3637.574 \\
  L1.M2SL                 & -32.2253 & 10.4440 & -3.085 & 0.002 & -52.695 & -11.755 \\
  L2.Total consumer credit & 0.1071 & 0.0410 & 2.596 & 0.009 & 0.026 & 0.188 \\
  L2.UMCSENT              & -4.1452 & 106.9670 & -0.039 & 0.969 & -213.797 & 205.507 \\
  L2.FEDFUNDS             & -98.6498 & 795.0990 & -0.124 & 0.901 & -1657.016 & 1459.716 \\
  L2.CPIAUCSL             & 548.6079 & 1029.4580 & 0.533 & 0.594 & -1469.093 & 2566.308 \\
  L2.M2SL                 & 41.2561 & 10.0550 & 4.103 & 0.000 & 21.548 & 60.964 \\
  \bottomrule
  \end{tabularx}
\end{table}

\begin{table}[H]
  \centering\scriptsize
  \caption{Lagged Parameters for Equation: UMCSENT}
  \label{tab:umcsent}
  \begin{tabularx}{\textwidth}{l*6{S[table-format=8.4]}}
  \LagTableHeader
  L1.Total consumer credit & 0.000000165 & 0.0000156 & 0.011 & 0.992 & -0.0000304 & 0.0000307 \\
  L1.UMCSENT              & -0.1034 & 0.0410 & -2.531 & 0.011 & -0.183 & -0.023 \\
  L1.FEDFUNDS             & -0.5521 & 0.2960 & -1.867 & 0.062 & -1.132 & 0.027 \\
  L1.CPIAUCSL             & -1.0467 & 0.3830 & -2.734 & 0.006 & -1.797 & -0.296 \\
  L1.M2SL                 & -0.0095 & 0.0040 & -2.439 & 0.015 & -0.017 & -0.002 \\
  L2.Total consumer credit & -0.000007675 & 0.0000155 & -0.496 & 0.620 & -0.000038 & 0.0000226 \\
  L2.UMCSENT              & -0.0580 & 0.0400 & -1.447 & 0.148 & -0.137 & 0.021 \\
  L2.FEDFUNDS             & -0.8939 & 0.2980 & -3.000 & 0.003 & -1.478 & -0.310 \\
  L2.CPIAUCSL             & 0.1828 & 0.3860 & 0.474 & 0.636 & -0.573 & 0.939 \\
  L2.M2SL                 & 0.0063 & 0.0040 & 1.675 & 0.094 & -0.001 & 0.014 \\
  \bottomrule
  \end{tabularx}
\end{table}

\begin{table}[H]
  \centering\scriptsize
  \caption{Lagged Parameters for Equation: FEDFUNDS}
  \label{tab:fedfunds}
  \begin{tabularx}{\textwidth}{l*6{S[table-format=8.4]}}
  \LagTableHeader
  L1.Total consumer credit & 0.000000186 & 0.00000203 & 0.091 & 0.927 & -0.00000379 & 0.00000416 \\
  L1.UMCSENT              & 0.0087 & 0.0050 & 1.638 & 0.101 & -0.002 & 0.019 \\
  L1.FEDFUNDS             & 0.4466 & 0.0380 & 11.606 & 0.000 & 0.371 & 0.522 \\
  L1.CPIAUCSL             & 0.0186 & 0.0500 & 0.373 & 0.709 & -0.079 & 0.116 \\
  L1.M2SL                 & -0.00006884 & 0.0010 & -0.135 & 0.893 & -0.001 & 0.001 \\
  L2.Total consumer credit & 0.000000231 & 0.00000201 & 0.115 & 0.909 & -0.00000371 & 0.00000418 \\
  L2.UMCSENT              & 0.0094 & 0.0050 & 1.792 & 0.073 & -0.001 & 0.020 \\
  L2.FEDFUNDS             & -0.2468 & 0.0390 & -6.362 & 0.000 & -0.323 & -0.171 \\
  L2.CPIAUCSL             & 0.0110 & 0.0500 & 0.220 & 0.826 & -0.087 & 0.109 \\
  L2.M2SL                 & -0.00008717 & 0.0000 & -0.178 & 0.859 & -0.001 & 0.001 \\
  \bottomrule
  \end{tabularx}
\end{table}

\begin{table}[H]
  \centering\scriptsize
  \caption{Lagged Parameters for Equation: CPIAUCSL}
  \label{tab:cpi}
  \begin{tabularx}{\textwidth}{l*6{S[table-format=8.4]}}
  \LagTableHeader
  L1.Total consumer credit & -0.000000226 & 0.00000164 & -0.138 & 0.891 & -0.00000344 & 0.00000299 \\
  L1.UMCSENT              & 0.0084 & 0.0040 & 1.954 & 0.051 & -0.000025 & 0.017 \\
  L1.FEDFUNDS             & 0.0463 & 0.0310 & 1.490 & 0.136 & -0.015 & 0.107 \\
  L1.CPIAUCSL             & 0.5126 & 0.0400 & 12.722 & 0.000 & 0.434 & 0.592 \\
  L1.M2SL                 & -0.0007 & 0.0000 & -1.777 & 0.076 & -0.002 & 0.0000753 \\
  L2.Total consumer credit & 0.000004303 & 0.00000163 & 2.645 & 0.008 & 0.00000111 & 0.00000749 \\
  L2.UMCSENT              & 0.0039 & 0.0040 & 0.924 & 0.355 & -0.004 & 0.012 \\
  L2.FEDFUNDS             & 0.0359 & 0.0310 & 1.144 & 0.253 & -0.026 & 0.097 \\
  L2.CPIAUCSL             & -0.1398 & 0.0410 & -3.443 & 0.001 & -0.219 & -0.060 \\
  L2.M2SL                 & 0.0016 & 0.0000 & 3.949 & 0.000 & 0.001 & 0.002 \\
  \bottomrule
  \end{tabularx}
\end{table}

\begin{table}[H]
  \centering\scriptsize
  \caption{Lagged Parameters for Equation: M2SL}
  \label{tab:m2sl}
  \begin{tabularx}{\textwidth}{l*6{S[table-format=8.4]}}
  \LagTableHeader
  L1.Total consumer credit & -0.00009044 & 0.0000 & -0.540 & 0.589 & -0.000 & 0.000 \\
  L1.UMCSENT              & -0.2536 & 0.4390 & -0.578 & 0.563 & -1.113 & 0.606 \\
  L1.FEDFUNDS             & -5.6777 & 3.1730 & -1.789 & 0.074 & -11.897 & 0.542 \\
  L1.CPIAUCSL             & -9.3560 & 4.1100 & -2.277 & 0.023 & -17.411 & -1.301 \\
  L1.M2SL                 & 0.8694 & 0.0420 & 20.693 & 0.000 & 0.787 & 0.952 \\
  L2.Total consumer credit & 0.0004 & 0.0000 & 2.435 & 0.015 & 0.0000788 & 0.001 \\
  L2.UMCSENT              & -0.0758 & 0.4300 & -0.176 & 0.860 & -0.919 & 0.768 \\
  L2.FEDFUNDS             & 0.8997 & 3.1990 & 0.281 & 0.778 & -5.369 & 7.169 \\
  L2.CPIAUCSL             & 3.1963 & 4.1410 & 0.772 & 0.440 & -4.921 & 11.313 \\
  L2.M2SL                 & -0.1897 & 0.0400 & -4.690 & 0.000 & -0.269 & -0.110 \\
  \bottomrule
  \end{tabularx}
\end{table}

\begin{longtable}{lS[table-format=8.4]S[table-format=8.4]S[table-format=8.4]S[table-format=8.4]S[table-format=8.4]}
\caption{Loading Coefficients ($\alpha$) for Each Equation}\label{tab:alpha}\\
\toprule
& \multicolumn{1}{c}{Total Credit} & \multicolumn{1}{c}{UMCSENT} & \multicolumn{1}{c}{FEDFUNDS} & \multicolumn{1}{c}{CPIAUCSL} & \multicolumn{1}{c}{M2SL}\\
\midrule
\endfirsthead
\toprule
& \multicolumn{1}{c}{Total Credit} & \multicolumn{1}{c}{UMCSENT} & \multicolumn{1}{c}{FEDFUNDS} & \multicolumn{1}{c}{CPIAUCSL} & \multicolumn{1}{c}{M2SL}\\
\midrule
\endhead
ec1 & -0.0015 & -0.0000003302 & 0.00000001058 & -0.0000001241 & -0.000003715 \\
ec2 & 137.1563 & -0.0193 & 0.0018 & -0.0048 & 0.2683 \\
\bottomrule
\end{longtable}

\begin{table}[H]
  \centering\scriptsize
  \caption{Cointegration Vector (Column 1)}
  \label{tab:cointeg1}
  \begin{tabularx}{\textwidth}{l*6{S[table-format=8.4]}}
  \toprule
  {Variables} & {Coef.} & {Std.\ Err.} & {$z$} & {$P>|z|$} & {95\% CI Lower} & {95\% CI Upper} \\
  \midrule
  Consumer Credit & 1.0000 & 0 & 0 & 0.000 & 1.000 & 1.000 \\
  Consumer Confidence & 1.89\text{e-}13 & 0 & 0 & 0.000 & 1.89\text{e-}13 & 1.89\text{e-}13 \\
  Fed Funds Rate & -35740 & 38500 & -0.928 & 0.353 & -1.11\text{e5} & 3.97\text{e4} \\
  CPI & 1793.4985 & 0.640 & 2804.145 & 0.000 & 1792.245 & 1794.752 \\
  M2 & -424.1162 & 3992.706 & -0.106 & 0.915 & -8249.677 & 7401.445 \\
  const & -1.35\text{e6} & 0.066 & -2.04\text{e7} & 0.000 & -1.35\text{e6} & -1.35\text{e6} \\
  \bottomrule
  \end{tabularx}
\end{table}

\begin{table}[H]
  \centering\scriptsize
  \caption{Cointegration Vector (Column 2)}
  \label{tab:cointeg2}
  \begin{tabularx}{\textwidth}{l*6{S[table-format=8.4]}}
  \toprule
  {Parameter} & {Coef.} & {Std.\ Err.} & {$z$} & {$P>|z|$} & {95\% CI Lower} & {95\% CI Upper} \\
  \midrule
  $\beta_1$ & 4.78\text{e-}20 & 0 & 0 & 0.000 & 4.78\text{e-}20 & 4.78\text{e-}20 \\
  $\beta_2$ & 1.0000 & 0 & 0 & 0.000 & 1.000 & 1.000 \\
  $\beta_3$ & -1.3650 & 53.276 & -0.026 & 0.980 & -105.785 & 103.055 \\
  $\beta_4$ & 0.0838 & 0.001 & 94.739 & 0.000 & 0.082 & 0.086 \\
  $\beta_5$ & 0.0012 & 5.31\text{e5} & 2.29\text{e-}09 & 1.000 & -1.04\text{e6} & 1.04\text{e6} \\
  const & -89.8497 & 8.820 & -10.187 & 0.000 & -107.136 & -72.564 \\
  \bottomrule
  \end{tabularx}
\end{table}

\end{document}